

\def\di{\partial}
\def\nab{\nabla}
\def\bnab{\nabla^*}
\def\lix{${\it \$}$_\xi}
\def\Lix{\hbox{{\it \$}}_\xi}
\def\de{\acute\partial}
\def\dez{\de^*}


\def\sig{\sigma}
\def\bsig{\sigma^*}
\def\bxi{{\bar\xi}}
\def\bXi{\bar\Xi}
\def\lam{\lambda}
\def\k{\kappa}
\def\alf{\alpha}
\def\ga{\gamma}


\def\gmn{g_{\mu\nu}}
\def\er{\vec e_r}
\def\half{{\textstyle{1 \over 2}}}
\def\jm{\hat j^\mu}
\def\Jm{\hat J^\mu}
\def\dom{{d\Omega \over 4\pi}}

\def\L{\vec L}
\def\P{\vec P}
\def\Z{\vec Z}
\def\M{{\cal M}}
\def\lag{\hat{\cal L}}

\def\m{\eqno}
\def\l{\left}
\def\r{\right}
\def\disp{\displaystyle}
\def\txt{\textstyle}
\def\intl{\int\limits}
\def\goto{\rightarrow}
\def\Bar{\overline}
\def\lft{\leftline}
\def\const{\rm const}
\def\SW{\hbox{\erm SW}}


\def\eit{\eightit}
\def\ebf{\eightbf}
\def\erm{\eightrm}
\font\eightrm=cmr8
\font\eightbf=cmbx8
\font\eightit=cmti8


\magnification = \magstep 1
\baselineskip = 22 pt

\centerline{\bf ON GLOBAL CONSERVATION LAWS AT NULL INFINITY}

\vskip .5 in

\centerline{Joseph Katz\footnote*{E-mail: JKATZ@VMS.HUJI.AC.IL}
            and Dorit Lerer}
\centerline{The Racah Institute of Physics, 91904 Jerusalem, Israel}

\vskip .5 in

9/12/1996

\centerline{ABSTRACT}

\vskip .4 in

  The ``standard'' expressions for total energy, linear momentum and also
angular momentum of asymptotically flat Bondi metrics at null infinity are also
obtained from differential conservation laws on asymptotically flat
backgrounds, derived from a quadratic Lagrangian density by methods currently
used in classical field theory. It is thus a matter of taste and commodity to
use or not to use a reference spacetime in defining these globally conserved
quantities. Backgrounds lead to  N\oe ther conserved currents; the use of
backgrounds is in line with classical views on conservation laws. Moreover, the
conserved quantities are in principle explicitly related to the sources of
gravity through Einstein's equations, while standard definitions are not.
The relations depend, however, on a rule for mapping spacetimes on backgrounds.

\filbreak

\beginsection 1. Introduction

  In {\sl Science and Hypothesis}, Poincar\'e (1904) imagines astronomers
  ``whose vision would be bounded by the solar system'' because of
  thick clouds that hide the fixed stars. There is thus no fixed frame of
  coordinates, only relative distances and relative angles are measurable. In
  those circumstances, says Poincar\'e, ``we should be definitively led to
  conclude that the equations which define distances are of an order higher
  than the second. [...] The values of the distances at any given moment
  depend upon their initial values, on that of their first derivatives, and
  something else. What is that {\it something else}? If we do not want it to be
  merely one of the second derivatives, we have only the choice of
  hypotheses. Suppose, as is usually done that this something else is the
  absolute orientation of the universe in space, or the rapidity with which
  this orientation varies; this may be, it certainly is, the most convenient
  solution for the geometer. But it is not the most satisfactory for the
  philosopher, because this orientation does not exist.''

  The conservation of energy, linear and angular-momentum, so useful
  in classical mechanics and special relativity, are related to the homogeneity
  and isotropy properties attributed to an absolute ``background'' spacetime
  which does indeed not
  exist. In general relativity, the use of backgrounds is intrinsic to the
  definition of pseudo-tensor conservation laws. Rosen (1958) and Cornish
  (1964) used backgrounds explicitly to calculate the energy-momentum in more
  appropriate coordinates than orthogonal ones. While backgrounds have thus
  proven to be useful, they nevertheless got little support and attention from
  the community of general relativists. On the contrary, great efforts have
  been spend to get rid of backgrounds, to avoid those ``additional structure
  completely counter spirit of general relativity'' (Wald 1984). True, efforts
  to avoid introducing background geometries
  generated interesting works notably by Penrose (1965), Geroch (1976) and
  other mathematical physicists mentioned in Schmidt's (1993) review on {\sl
  Asymptotic flatness --- a Critical Appraisal}. As a consequence, we now have
  a rather well understood, coordinate independant, picture of asymptotic
  flatness, and we have also asymptotic coordinate independant expressions for
  globally conserved quantities in particular at null infinity.

  Conserved quantities at null infinity are given by integrals on spheres of
  infinite radius at fixed null time $u=t-r=\const$. They represent total
  energy, linear momentum, angular-momentum and the initial position of the
  mass center (Synge 1964)
  of spacetimes with isolated sources of curvature at a moment of ``time''
  $u$. The expressions have become ``standard'' according to Dray and
  Streubel (1984) [see also Dray (1985) and Shaw (1986)], and they are
  invariant under coordinate transformations of the Bondi-Metzner-Sachs, or
  BMS, group [Sachs (1962), Newman and Penrose (1966)]. Most standard
  expressions draw their strength --- beyond their esthetical appeal --- from a
  few physically interpretable quantities, which have all been obtained before
  from the pseudo-tensors of Freud (1939) and of Landau and Lifshitz
  (1951). The quantities are
 \item {(a)} The Schwarzschild mass $M_{\rm Schw}$ and more generally the
             dominant $1/r$
             term in a multipole expansion of static solutions at spatial
             infinity (Geroch 1970) and at null infinity (Schmidt 1993), which
             was obtained from Einstein's pseudo-tensor [see Tolman 1934].
 \item {(b)} The Bondi mass $M(u)$ where $u=t-r$ [see Bondi, Metzner and van
             der Burg (1962) see already Bondi (1960)] and Sachs's (1962a)
             linear momentum $\P(u)$ at null infinity\footnote*{What is
             actually defined in Sachs is $d\P/du$.} which have been calculated
             by M\o ller (1972) using Freud's superpotential;
 \item {(c)} The Kerr angular-momentum $\L_{\rm Kerr}$ and more generally the
             dominant
             $1/r^2$ ``odd function'' factor [Misner Thorne and Wheeler (1973)]
             in a multipole expansion of stationary spacetimes at spatial or
             null infinity, which is obtainable from Landau and Lifshitz
             superpotential for angular momentum [Papapetrou 1965].

  The angular momentum at null infinity $\L(u)$, as given for instance by Dray
  and Streubel (1984), has not been derived from previously known
  soperpotential. $\L(u)$ is not a well defined physical quantity because
  of supertranslation freedom. However, the weak field approximation fits
  in well with results deduced from Landau-Lifshitz's superpotential
  (Creswell and Zimmerman 1986). The LL approximation is currently used in
  gravitational radiation calculations (Thorne 1980).

    Here we show that $M(u)$, $\P(u)$ as well as $\L(u)$ --- and thus also
  $M_{\rm Schw}$ and $\L_{\rm Kerr}$ --- can be calculated using a single
  superpotential derived from a quadratic Lagrangian with an asymptotically
  flat background at null infinity. The derivation insures automatically
  Poincar\'e invariance, the absence of an anomalous factor of 2 in the $M/L$
  ratio and zero value for all conserved quantities when the spacetime
  identifies with the background itself. As a result, $dM/du$ and $d\P/du$ are
  the standard BMS invariant fluxes while $d\L/du$, which is not BMS invariant,
  is nevertheless the same as the standard result. The position of the mass
  center, $\Z(u)$, associated with Lorentz rotations, and the corresponding
  flux $d\Z/du$ given here are not the same as the standard formula, at least
  in the non-linear approximation.

  Thus, the most important standard results, energy and angular momentum,
  can be deduced from N\oe ther conservation laws. One appealing aspect of N\oe
  ther conservation laws is
  that they are closer to classical intuition than the more abstract coordinate
  independant definitions at scri. Moreover, N\oe ther conserved quantities are
  directly related to the energy-momentum tensor of the matter through
  Einstein's equations by differential conservation laws. ``Standard''
  definitions are not.

  Local differential conservation laws contain implicit
  definitions of local or quasi-local conserved quantities which depend, on the
  {\sl local} mapping of the spacetime on the background. We give here no rule
  for local mappings. There are also several mapping-independent
  definitions of quasi-local energy in the literature, but they appear to
  be different from each other (Berqvist 1992).

\beginsection 2. Superpotential and N\oe ther Conservation Laws

  Let us briefly review the elements that lead to our superpotential (Katz
  1996). Full details are given in a work by Katz, Bi\u c\'ak and Lynden-Bell
  (1996) --- refered to below as KBL96 --- which has its roots in earlier work
  by Katz (1985) on flat backgrounds. KBL96 is on curved backgrounds and the
  superpotential obtained there is new {\sl even in the limit where the
  backgrounds become flat}.
\nobreak
\beginsection {\it (i) Lagrangian density for gravitational fields on a
                             curved background}

  Let $g_{\mu\nu}(x^\lam),~~ \lam,\mu,\nu,...= 0,1,2,3$ be the metric of a
  spacetime $\M$ with signature -2, and let $\bar g_{\mu\nu}({\bar x}^\lam)$ be
  the metric of the background $\bar\M$. Both are tensors with respect to
  arbitrary coordinate transformations. Once we have chosen a mapping so that
  points $P$ of $\M$ map into points $\bar P$ of $\bar\M$,
  then we can use the convention that $\bar P$ and $P$ shall always be given
  the same coordinates ${\bar x}^\lam=x^\lam$. This convention implies that a
  coordinate transformation on $\M$ inevitably induces a coordinate
  transformation with the same functions on $\bar\M$ . With this convention,
  such expressions as $g_{\mu\nu}(x^\lam)-\bar g_{\mu\nu}(x^\lam)$, which can
  be looked at as ``perturbations'' of the background, become true
  tensors. However, if the particular mapping has been left unspecified we are
  still free to change it. The form of the equations for ``perturbations'' of
  the background must inevitably contain a gauge invariance corresponding to
  this freedom.

  Let $R^\lam_{~\nu\rho\sig}$ and ${\bar R}^\lam_{~\nu\rho\sig}$ be the
  curvature tensors of $\M$ and $\bar\M$. These are related as follows [see
  Rosen (1940, 1963), see also Choquet-Bruhat (1984) for mathematical
  aspects of background formalisms]:
 $$R^\lam_{~\nu\rho\sig}=\bar D_\rho\Delta^\lam_{\nu\sig}
                          -\bar D_\sig\Delta^\lam_{\nu\rho}
                          +\Delta^\lam_{\rho\eta}\Delta^\eta_{\nu\sig}
                          -\Delta^\lam_{\sig\eta}\Delta^\eta_{\nu\rho}
                          +{\bar R}^\lam_{~\nu\rho\sig}.              \m[2-1]$$
  Here $\bar D_\rho$ are covariant derivatives with respect to $\bar
  g_{\mu\nu}$, and $\Delta^\lam_{\mu\nu}$ is the difference between Christoffel
  symbols in $\M$ and $\bar\M$:
   $$\Delta^\lam_{\mu\nu}=\Gamma^\lam_{\mu\nu}-{\bar\Gamma}^\lam_{\mu\nu}
                         =\half g^{\lam\rho}(\bar D_\mu g_{\rho\nu}
                                            +\bar D_\nu g_{\rho\mu}
                                            -\bar D_\rho g_{\mu\nu}).\m[2-2]$$

  Our quadratic Lagrangian density $\lag_G$ for the gravitational field is then
  defined as
   $$\lag_G=\lag-\Bar{\lag}~~~,~~~
     \lag=-{1\over 2\k}(\hat R+\di_\mu{\hat k}^\mu)~~~,~~~
     \Bar{\lag}=-{1\over 2\k}\Bar{\hat R}~~~,~~~\k={8\pi G\over c^4}.\m[2-3]$$
  The mark $~\hat{ }~$ means multiplication by $\sqrt{-g}$, never by
  $\sqrt{-\bar g}$, and a bar over a symbol signifies that $\gmn$, $D_\rho$
  etc. are replaced by $\bar\gmn$, $\bar D_\rho$
  etc.. The vector density ${\hat k}^\mu$ is given by
   $${\hat k}^\mu={1\over\sqrt{-g}}\bar D_\nu(-gg^{\mu\nu})
                 ={\hat g}^{\mu\rho}\Delta^\sig_{\rho\sig}
                  -{\hat g}^{\rho\sig}\Delta^\mu_{\rho\sig},         \m[2-4]$$
  and its divergence cancels all second order derivatives of $g_{\mu\nu}$ in
  $R$. $\lag$ is the
  Lagrangian used by Rosen. $\Bar{\lag}$ is $\lag$ in which $g_{\mu\nu}$ has
  been replaced by $\bar g_{\mu\nu}$. When $g_{\mu\nu}=\bar g_{\mu\nu}$,
  $\lag_G$ is thus identically zero. The intention here is to obtain
  conservation laws in the background spacetime so that if $g_{\mu\nu}=\bar
  g_{\mu\nu}$, conserved vectors and superpotentials will be identically zero
  as in Minkowski space in special relativity.

  The following formula, deduced from [2-3] and [2-1],
  shows explicitly how $\lag_G$ is quadratic in the first order derivatives of
  $g_{\mu\nu}$ or, equivalently, quadratic in $\Delta^\mu_{\rho\sig}$:
   $$\lag_G={1\over 2\k}{\hat g}^{\mu\nu}
                 (\Delta^\rho_{\mu\nu}\Delta^\sig_{\rho\sig}
                   -\Delta^\rho_{\mu\sig}\Delta^\sig_{\rho\nu})
            -{1\over 2\k}({\hat g}^{\mu\nu}-\Bar{{\hat g}^{\mu\nu}})
                                                    \bar R_{\mu\nu}. \m[2-5]$$
  Notice that if ${\bar R}^\lam_{~\nu\rho\sig}=0$ and coordinates are such that
  ${\bar \Gamma}^\lam_{\mu\nu}=0$, $\lag_G$ is nothing else than the familiar
  ``$\Gamma\Gamma-\Gamma\Gamma$'' Lagrangian density [see for instance Landau
  and Lifshitz (1951)].

\beginsection {\it (ii) Strong conservation laws and superpotential}

  If	
	$$\Delta x^\mu=\xi^\mu\Delta\lam	\m[2-6]$$
  represents an infinitesimal one parameter displacement generated by
  $\xi^\mu$, the corresponding changes in tensors are given in terms of the Lie
  derivatives with respect to the vector field $\xi^\mu$, $\Delta
  g_{\mu\nu}=\lix g_{\mu\nu}\Delta\lam$, etc.. The Lie derivatives may be
  written in terms of ordinary partial derivatives $\di_\mu$, covariant
  derivatives $\bar D_\mu$ with respect to $\bar g_{\mu\nu}$, or covariant
  derivatives $D_\mu$ with respect to $g_{\mu\nu}$. Thus,
   $$\eqalignno{ \Lix\gmn
            &=g_{\mu\lam}\di_\nu\xi^\lam+g_{\nu\lam}\di_\mu\xi^\lam
                                             +\xi^\lam\di_\lam\gmn   &[2-7a]\cr
            &=g_{\mu\lam}\bar D_\nu\xi^\lam+g_{\nu\lam}\bar D_\mu\xi^\lam
                                            +\xi^\lam\bar D_\lam\gmn &[2-7b]\cr
            &=g_{\mu\lam} D_\nu\xi^\lam+g_{\nu\lam} D_\mu\xi^\lam.&[2-7c]\cr}$$

  Consider now the Lie differential $\Delta\lag$ of $\lag$. With the
  variational principle in mind, we write $\Delta\lag=\lix\lag\Delta\lam$ in
  the form
   $$\Delta\lag=\txt{1\over 2\k}\hat G^{\mu\nu}\Delta\gmn
                +\di_\mu\hat A^\mu\Delta\lam                    \m[2-8]$$
  where Einstein's tensor density,
  $\hat G^{\mu\nu}=\hat R^{\mu\nu}-\half\hat g^{\mu\nu}R$, is the variational
  derivative of $\lag$ with respect to $\gmn$, $\hat A^\mu$ is a vector density
  linear in $\xi^\mu$ whose detailed form will not concern us here. The Lie
  derivative of a scalar density like $\lag$ is just an ordinary divergence
  $\di_\mu(\lag\xi^\mu)$, Thus
   $$\hat O\equiv\Lix\lag-\di_\mu(\lag\xi^\mu)=0.                \m[2-9]$$
  Combining [2-9] with [2-8], we obtain
   $$\hat O\equiv \txt{1\over 2\k}\hat G^{\mu\nu}\Lix\gmn
                  +\di_\mu(\hat A^\mu-\lag\xi^\mu).              \m[2-10]$$

  Bianchi identities imply $D_\nu G^{\mu\nu}=0$ so that with [2-7c], [2-10] can
  be written as the divergence of a vector density $\jm$, say,
   $$\hat O=\di_\mu\jm=0~~~~~~~~{\rm where}~~~~~~~
     \jm=\txt{1\over\k}\hat G^\mu_\nu\xi^\nu+\hat B^\mu.        \m[2-11]$$
  Hence, $\lag$  ``generates'' a vector density $\jm$ that is {\sl identically}
  conserved. It has been obtained without using Einstein's field equations;
  [2-11] is the kind of strong conservation law introduced by Bergmann
  (1949). We shall, of course, assume that Einstein's equations are satisfied,
  and replace $\txt{1\over\k} G^\mu_\nu$ by the energy-momentum of matter
   $$	\txt{1\over\k} G^\mu_\nu=T^\mu_\nu                       \m[2-12]$$
  so that our strong conservation law [2-11] reads:
   $$\di_\mu\jm=\di_\mu(\hat T^\mu_\nu \xi^\nu + \hat B^\mu)=0.  \m[2-13]$$
  Equations [2-13] are, strictly speaking, not identities anymore. Given
  $T^\mu_\nu$, [2-13] holds only for metrics that satisfy [2-12]. The vector
  density $\jm$ is linear in $\xi^\mu$ and its derivatives up to order 2.

  Since $\jm$ as given by  [2-11] is identically conserved whatever is $\gmn$,
  it must be the divergence of an antisymmetric tensor density that depends
  on the arbitrary $\gmn$'s as well;  thus
   $$\jm = \di_\nu\hat j^{\mu\nu},~~~~~~~~{\rm where}~~~~~~~
     \hat j^{\mu\nu}=-\hat j^{\nu\mu}.                          \m[2-14]$$
  Indeed, $\hat j^{\mu\nu}$ is easy to find and is derived directly from $\lag$
  in Katz\footnote*{In the 1985 paper the background is assumed to be flat,
  but the derivation of $\hat j^{\mu\nu}$ does not depend on that assumption.}
  (1985) [see also Chru\'sciel (1986), Sorkin (1988) and Katz and Ori (1990)]:
   $$ j^{\mu\nu}=\txt{1\over\k}D^{[\mu}\xi^{\nu]}
                +\txt{1\over\k}\xi^{[\mu}k^{\nu]}.              \m[2-15]$$
  The terms $\txt{1\over\k}D^{[\mu}\xi^{\nu]}$ will be recognised as $\half$
  Komar's(1959) superpotential. In terms of $\bar D_\rho$ derivatives,
   $$D_\rho\xi^\mu=\bar D_\rho\xi^\mu+\Delta^\mu_{\rho\lam}\xi^\lam, \m[2-16]$$
  and, using expression [2-4] for $k^\mu$, $j^{\mu\nu}$ may be written in
  the form
   $$\k j^{\mu\nu}=g^{[\mu\rho}\bar D_\rho\xi^{\nu]}
                  +g^{[\mu\rho}\Delta^{\nu]}_{\rho\lam}\xi^\lam
                  +\xi^{[\mu}g^{\nu]\rho}\Delta^\sig_{\rho\sig}
                  -\xi^{[\mu}\Delta^{\nu]}_{\rho\sig}g^{\rho\sig},  \m[2-17]$$

  Had we applied the identities [2-9] to $\Bar{\lag}$ instead of $\lag$, we
  would have written everywhere $\Bar\gmn$ instead of $\gmn$. We would have
  found strong, barred, conserved vector densities $\Bar{\jm}$ and barred
  superpotentials $\Bar{\hat j^{\mu\nu}}$ with the same $\xi^\mu$'s:	
   $$\Bar{\jm}=\Bar{\hat T^\mu_\nu}\xi^\nu+\Bar{\hat B^\mu}
              =\di_\nu \Bar{\hat j^{\mu\nu}},                      \m[2-18]$$
  with
  $$\Bar{\hat j^{\mu\nu}} = \txt{1\over\k}\Bar{D^{[\mu}\hat\xi^{\nu]}}
                            ~~~~~( \Bar{\hat k^\mu}\equiv 0 ).     \m[2-19]$$

  Strongly conserved vectors for $\lag_G=\lag-\Bar{\lag}$ are obtained by
  subtracting barred vectors and superpotentials from unbarred ones; in this
  way we define {\sl relative} vectors and in particular {\sl relative
  superpotentials} $\hat J^{\mu\nu}$ --- relative to the background
  space. Setting
   $$\hat I^\mu=\jm-\Bar{\jm}~~~~~~~,~~~~~~~
     \hat J^{\mu\nu}=\hat j^{\mu\nu}-\Bar{\hat j^{\mu\nu}}
                    =-\hat J^{\nu\mu},                             \m[2-20]$$
  we have for the strongly conserved vector $\hat I^\mu$ the following form:
   $$\hat I^\mu = \hat J^\mu + \hat\zeta^\mu
                =\di_\nu \hat J^{\mu\nu},~~~~{\rm and}~~~~~~
     \di_\mu\hat I^\mu \equiv 0,                                   \m[2-21]$$
  which hold for any $\xi^\mu$ and any mapping of $\M$ on $\bar\M$;
  in [2-21]
   $$\hat J^\mu=\hat\theta^\mu_\nu \xi^\nu
                +\hat\sig^{\mu[\rho\sig]}\di_{[\rho}\xi_{\sig]}.   \m[2-22]$$
  $\hat\theta^\mu_\nu$, $\hat\sig^{\mu\rho\sig}$ and $\hat\zeta^\mu$ are given
  explicitly in appendix for the interested reader.
  The relative superpotential density $\hat J^{\mu\nu}$ is now given by
   $$\hat J^{\mu\nu}=\txt{1\over\k}( D^{[\mu}\hat\xi^{\nu]}
                                    -\Bar{D^{[\mu}\hat\xi^{\nu]}}
                                    +\hat\xi^{[\mu}k^{\nu]} ),     \m[2-23]$$
  and can be also written in terms of $\gmn$, $\Delta^\mu_{\rho\sig}$ and
  $\xi^\mu$:
   $$\k\hat J^{\mu\nu} =\hat l^{[\mu\rho}\bar D_\rho\xi^{\nu]}
                 +g^{[\mu\rho}\Delta^{\nu]}_{\rho\lam}\hat\xi^\lam
                 +\hat\xi^{[\mu}g^{\nu]\rho}\Delta^\sig_{\rho\sig}
                 -\hat\xi^{[\mu}\Delta^{\nu]}_{\rho\sig}g^{\rho\sig},\m[2-24]$$
  in which
   $$\hat l^{\mu\nu}=\hat g^{\mu\nu}-\Bar{\hat g^{\mu\nu}}.          \m[2-25]$$

  The tensors in [2-22] have a physical interpretation. On a flat background,
  in coordinates in which $\bar\Gamma^\lam_{\mu\nu}=0$, [see appendix]
   $$\hat\theta^\mu_\nu=\hat T^\mu_\nu + \hat t^\mu_\nu ,           \m[2-26]$$
  and $\hat t^\mu_\nu$ reduces to Einstein's pseudo-tensor density.
  $\hat\theta^\mu_\nu$ appears therefore as the energy-momentum tensor of the
  gravitational field with respect to the background. The second tensor in
  [2-22], $\hat\sig^{\mu[\rho\sig]}$, is quadratic in the metric perturbations
  just like $t^\mu_\nu$. It is also bilinear in the perturbed metric components
  ($\gmn-\bar\gmn$) and their first order derivatives.
  $\hat\sig^{\mu[\rho\sig]}$ resembles, in this respect, the helicity tensor
  density in electromagnetism. The factor of $\di_{[\rho}\xi_{\sig]}$
  represents thus the helicity tensor density of the perturbations with respect
  to the background.

  It should be noted again that all the components of $I^\mu$ and of the
  superpotential $J^{\mu\nu}$ itself are identically zero if $\gmn=\bar\gmn$;
  therefore strong conservation laws refer to ``perturbations'' only and not to
  the background.

\beginsection {\it (iii) N\oe ther conservation laws}

  We now consider what happens when arbitrary $\xi^\mu$'s are replaced by
  Killing vectors $\bar\xi^\mu$  of the background. $\Jm$ defined in [2-22],
  which contains the physics of the conservation laws [see KLB96], is not, in
  general, a conserved vector density since the identically conserved vector
  density is $\hat I^\mu = \hat J^\mu + \hat\zeta^\mu$  and thus
   $$\di_\mu\Jm = - \di_\mu\hat\zeta^\mu.                        \m[2-27]$$
  However, when $\xi^\mu$ is a Killing vector of the background, $\bar\xi^\mu$,
  then [see appendix] $\hat\zeta^\mu = 0$ and $\Jm(\bar\xi)$ {\sl is}
  conserved.

  Our $\Jm$ has been derived in the same way as ``N\oe ther's theorem'' in
  classical field theory [see for instance Schweber, Bethe and Hoffmann
  (1956), or Bogoliubov and Shirkov (1959)].
  Thus, by replacing $\xi^\mu$ in strongly conserved
  currents by Killing vectors, $\bar\xi^\mu$, of the background we obtain N\oe
  ther conserved vector densities in general relativity with mappings on curved
  backgrounds.
   $$\Jm(\bar\xi)=\hat\theta^\mu_\nu \bar\xi^\nu
                   +\hat\sig^{\mu[\rho\sig]} \di_{[\rho}\bar\xi_{\sig]}
                 =\di_\nu\hat J^{\mu\nu}(\bar\xi), ~~~~~~
                    \di_\mu\Jm(\bar\xi)=0                           \m[2-28]$$
  with $\hat J^{\mu\nu}(\bar\xi)$ given by
   $$\k\hat J^{\mu\nu}(\bar\xi)=
           \hat l^{[\mu\rho} \bar D_\rho \bar\xi^{\nu]}
          +\hat g^{[\mu\rho}\Delta^{\nu]}_{\rho\lam} \bar\xi^{\lam}
          +\bar\xi^{[\mu} \hat g^{\nu]\rho} \Delta^{\sig}_{\rho\sig}
          -\bar\xi^{[\mu} \Delta^{\nu]}_{\rho\sig} \hat g^{\rho\sig}.\m[2-29]$$

  We can now integrate [2-28], on a part $\Sigma$ of a hypersurface $\cal S$,
  which spans a two-surface $\di\Sigma$, and obtain integral conservation
  laws:
   $${c^4\over G}{\cal P}(\bxi)
     = \intl_\Sigma  (\hat\theta^\mu_\nu \bxi^\nu
                +\hat\sig^{\mu[\rho\sig]} \di_{[\rho}\bxi_{\sig]}) d\Sigma_\mu
     = \intl_{\di\Sigma} \hat J^{\mu\nu}(\bar\xi) d\Sigma_{\mu\nu}.  \m[2-30]$$

  ${\cal P}(\bxi)$ depends only on the gravitational field and its first
  derivatives on $\di\Sigma$ and on the mapping near the boundary. The relation
  with the matter tensor depends, however, on the mapping all the way down to
  the sources of gravity.

  For weak fields on a flat background, the lowest order linear approximation
  of [2-30] is
   $${c^4\over G}{\cal P}(\bxi)
     = \intl_\Sigma \hat T^\mu_\nu \bxi^\nu
     = \intl_{\di\Sigma} \hat J^{\mu\nu}(\bar\xi) d\Sigma_{\mu\nu}.  \m[2-31]$$

  If the spacetime is asymptotically flat, intergrals over the whole
  hypersurface $\cal S$ extending to infinity define globally conserved
  quantities. It is then appropriate to map the spacetime near infinity on a
  flat Minkowski space with its ten Killing vectors
  associated with spacetime translations, spatial rotations and Lorentz
  rotations. The ten Killing vectors give ten different expressions ${\cal
  P}(\bxi)$, which can be interpreted respectively as the total energy $E$, the
  linear momentum vector $\P$, the angular momentum vector $\L$ and the initial
  mass center position $\Z$ on $\cal S$.

  We shall now calculate the conserved quantities for Bondi's
  asymptotic solution on a null hypersurface at infinity, using the right hand
  side of [2-30]. In what follows we intend to define all the quantities
  introduced. We shall not use cross-referencing for definitions which are
  often given with different symbols, factors of 2 or 1/$\sqrt{2}$ and other
  signs; most readers will appreciate this affort.

\bigskip

\vbox{
\beginsection 3. Elements of the Asymptotic BMS Metric in Newman-Unti
                 Coordinates

\beginsection {\it (i) The Newman-Unti asymptotic solution}

}
  In the coordinates used by Newman and Unti (1962) $x^\lam=(x^0=u , x^1=r ,
  x^2 , x^3)$, the metric of  Bondi, Metzner and Sachs (1962) has the following
  form:
  $$ds^2=g_{00}du^2+2dudr+2g_{0L}dudx^L+g_{KL}dx^Kdx^L~~~,~~~~K,L=2,3.\m[3-1]$$
  The metric components are given by Newman and Unti in their formula (41).
  We shall make a few changes of notations because several indices which make
  sense in Newman and Unti are not useful here. Thus
 \item * The indices of the metric where shifted from $\lam=1,2,3,4$ to
         $\lam=0,1,2,3$ ($\lam$ or any other lower-case greek letter) so that
         $x^1_{NU}=x^0=u$, $x^2_{NU}=x^1=r$, $x^3_{NU}=x^2$ and $x^4_{NU}=x^3.$
 \item * We have denoted real parts with a prime like $a'$, instead of Re$(a)$
         in NU, and imaginary parts with a ``second'' like $a''$, instead of
         Im$(a)$. Thus $\sig={\sig}'+i{\sig}''$ etc...
 \item * Complex conjugation is denoted here by an asterisk $a^*$, because a
         bar over the symbol is reserved for the background.
         Thus $\bar a_{NU}=a^*$.
 \item * The $\psi_j^\circ$ and $\sig^\circ$ of NU are here written $\psi_j$
         and $\sig$. Notice that $\psi_j$ and $\sig$ exist also in NU with a
         different meaning.
 \item * $O(r^{-n})$ of NU is here written $O_n$.

  With these changes of notations, the metric components given by Newman and
  Unti become as follows:
  $$\disp{\eqalign{
    g^{00} &=0~~~~g^{01}=1~~~~g^{02}=g^{03}=0   \cr
    g^{11} &=-2P^2\nab\bnab \ln P -{2{\psi_2}' \over r}
              +{ {\txt{2\over 3}}P^2[\nabla(\psi^*_1/ P)]'-\half(|f|^2/P^2)
                                                        \over r^2} +O_3 \cr
    g^{12}+ig^{13} &=-{f^* \over r^2}
                   +{{( {\txt{4\over 3}} P\psi_1+2\sig f)} \over r^3} +O_4 \cr
    g^{23} &={4P^2\sig'' \over r^3} + {4P^2|\sig|^2\sig'' \over r^5}  +O_6 \cr
    g^{22} &=-{2P^2\over r^2} + {4P^2\sig' \over r^3}
                               - {6 |\sig|^2 P^2 \over r^4}      +O_5   \cr
    g^{33} &=-{2P^2\over r^2} - {4P^2\sig' \over r^3}
                               - {6|\sig|^2 P^2 \over r^4} +O_5. \cr}}\m[3-2]$$
  $P$ is a function of $x^2$, $x^3$ while $\psi_1$, $\psi_2$ and $\sig$ are
  complex scalar functions of $u$, $x^2$, $x^3$, defined in terms of the null
  tetrad components of the Weyl tensor.
  $f$ is defined in terms of $\sig$ and $P$
   $$f\equiv 2P^4\nabla(\sig^*/P^2),                                 \m[3-3]$$
  and
     $$    \nabla\equiv\di_2+i\di_3.                                 \m[3-4]$$
  Note that we have expanded $g^{23}$ to the fifth order of $1/r$.

  From [3-2] we have calculated the $\gmn$ components:
  $$\disp{\eqalign{
    g_{10}        &=1~~~~~~g_{11}=g_{12}=g_{13}=0                 \cr
    g_{00}        &=2P^2\nab\bnab\ln P +\l[ {2\psi_2 \over r}
                   -{{{\txt{2\over 3}}P^2\nabla(\psi^*_1/P)}\over r^2}
                   +{1\over r^3}{{\txt{4\over 3}}fP\psi_1+\sig ff\over P^2}\r]'
                   +O_4  \cr
    g_{02}+ig_{03}&={1\over 2P^2}\left[-f^* +{{\txt{4\over 3}}P\psi_1 \over r}
                   +{{\txt{8\over 3}}P\sig\psi^*_1+3|\sig|^2 f^* \over r^2}
                   +O_3 \right]   \cr
    g_{23}     &=-r{\sig''\over P^2}+{1\over r}{|\sig|^2\sig''\over P^2}+O_2\cr
    g_{22}     &= - {r^2 \over 2P^2} - {r\sig' \over P^2}
                                     - {|\sig|^2 \over 2P^2}           +O_1 \cr
    g_{33}     &= - {r^2 \over 2P^2} + {r\sig' \over P^2}
                                  - {|\sig|^2 \over 2P^2} + O_1. \cr}}\m[3-5]$$

  A useful quantity in our calculation is the density
   $$\sqrt{-g}={r^2\over 2P^2}\left(1-{|\sig|^2 \over r^2}+O_3\right).\m[3-6]$$
  Further useful informations taken from Newman and Unti are
 \item {(a)} The $u$-derivative equations derived from the Bianchi identities,
             in the form given in their equations (40k,l) or (42b,c). We shall
             most of the time use a dot on a symbol to denote a derivative of
             this function with respect to $x^0=u$; thus
             $\dot\psi\equiv\di_0\psi$. With this change of notation the
             formulas are
     $$  \dot{\psi_1} - P\nab\psi_2 - 2\sig\psi_3 = 0                 \m[3-6]$$
     $$\dot{\psi_2} + \sig\ddot\bsig - P^2\nab\l(\psi_3 \over P\r)=0  \m[3-7]$$
   where
     $$\psi_3 = -P^3\nab
                  \l({\dot\bsig\over P^2} + {1\over P}\bnab\bnab P\r).\m[3-8]$$
 \item {(b)} The metric keeps the same form under coordinate transformations of
             the Bondi-Metzner-Sachs, or BMS, group (Sachs 1962b, Newman and
             Penrose 1966). The leading terms in powers of $1/r$ of the
             transformation $x^\lam \goto \tilde{x^\lam}$ are given by NU in
             their eq.~(46):
            $$\eqalignno{ \tilde u &= J(x^L)u + K(x^L)   +O_1    &[3-9] \cr
                          \tilde r &= {1 \over J}r       +O_0    &[3-10]\cr
                        \tilde x^K &= Y^K(x^L)           +O_1.   &[3-11]\cr }$$
             $J$ and $K$ are arbitrary functions and the $Y^K$ induce conformal
             transformations in $(x^K)$ space, i.e.
            $$\di_2 Y^2=\pm\di_3 Y^3~~~~,~~~~\di_3 Y^2=\mp\di_2 Y^3. \m[3-12]$$

  From this follows that $P(x^L)$ can be fixed with an appropriate conformal
  transformation [3-12], and $r$ can be fixed by choosing a spherical
  boundary for the surface $u=\const$. But $u$ is only defined up to a
  supertranslation $K(x^L)$. For a fixed $P$ and a fixed $r$ there are five
  independent scalar functions in the metric:
  ${\psi_1}'$ and ${\psi_1}''$, $\sig'$ and $\sig''$ and ${\psi_2}'$. But they
  are defined up to a supertranslation $K(x^L)$ and therefore there are
  actually four independent initial quantities among the ten $\gmn$'s on
  $u=u_0$. The imaginary part of $\psi_2$,  ${\psi_2}''$, is not independent;
  It is defined in terms of $P$ and $\sig$ [NU (40g)]:
     $$\psi_2''=\l(P^2\bnab{f^*\over 2P^2} + \bsig\dot\sig\r)''.\m[3-13]$$
  The physical interpretation of these functions has been analyzed in
  details in a series of papers by Bondi, Metzner and Sachs [see especially
  Bondi van der Burg and Metzner (VII) 1962]

\beginsection 4. Elements of the Asymptotic Background

  The asymptotic background is flat. In Minkowski coordinates
  $X^\alf=(X^0=t,X^k)$ $k,l,...=1,2,3$, the metric element
 $$d\bar s^2=\eta_{\alf\beta}dX^\alf dX^\beta=dt^2-d\vec X^2.        \m[4-1]$$
  An arrow designates spatial 3-vectors in Minkowski space. In $X^\alf$
  coordinates, the Killing vector components of the ten translations are given
  by
 $$\tilde\xi^\mu_\alf = \delta^\mu_\alf                              \m[4-2]$$
  and of rotations by
 $$\tilde\xi^\mu_{[\alf\beta]}=
            (\tilde\xi^\mu_\alf \eta_{\beta\ga}
            -\tilde\xi^\mu_\beta \eta_{\alf\ga}) X^\ga.              \m[4-3]$$

  In NU-coordinates $x^\lam=(u,r,x^K)$
 $$d\bar s^2=\bar g_{\alf\beta}dx^\alf dx^\beta
            =du^2+2dudr-{r^2\over 2P^2}\l[(dx^2)^2+(dx^3)^2\r],      \m[4-4]$$
  where
 $$r=\sqrt{{\vec X}^2}                                               \m[4-5]$$
  and the metric of a sphere ($u=u_0, r=r_0)$ in conformal coordinates has the
  well known Riemann form for spaces of constant curvature [Eisenhart 1922]:
 $$P=\half+\txt{1\over 4}[(x^2)^2+(x^3)^2].                          \m[4-6]$$
  With $P$ given by [4-6] we find that
 $$2P^2\nab\bnab\ln P=1.                                             \m[4-7]$$
  Eq. [4-7] somewhat simplifies $g^{11}$ given in [3-2]. Moreover, since $P$
  satisfies the following equation
 $$\bnab\bnab P  = 0,                                                \m[4-8]$$
  $\psi_3$ defined in [3-8] becomes also simpler:
 $$\psi_3=-P^3\nab\l(\dot\bsig\over P^2 \r)=-{\dot f \over 2P},      \m[4-9]$$
  $f$ has been defined in [3-3].

  $x^2$ and $x^3$ are not uniquely defined nor are they uniquely related to the
  spherical coordinates $(r,\theta,\phi)$ in $\bar\M$. By definition
  $X^1=r\sin\theta\cos\phi$, $X^2=r\sin\theta\sin\phi$ and $X^3=r\cos\theta$.
  We shall define $x^\mu(X^\lam)$ as follows:
 $$\eqalignno{
     x^0&=u=t-r ~~~~,~~~~ x^1=r                                 &[4-10] \cr
    x^2&=\sqrt{2}{P\over r}X^1=\sqrt{2}\cot(\half\theta)\cos\phi &[4-11] \cr
   x^3&=-\sqrt{2}{P\over r}X^2=-\sqrt{2}\cot(\half\theta)\sin\phi.&[4-12]\cr}$$
  This choice makes the connection with the formalism of Newman and Penrose
  (1966) simple, as we shall see below. In terms of spherical coordinates that
  are sometimes useful in the calculations, $P$ becomes
 $$P={1 \over 2\sin^2(\half\theta)}.                                \m[4-13]$$

  Let us introduce a unit vector in the radial direction in Minkowski space,
  denoted by $\er$,
 $$\er = {\vec X \over r}
       = (\sin\theta\cos\phi,~\sin\theta\sin\phi,~\cos\theta),      \m[4-14]$$
  and the complex 2-vector $\bXi$ on the sphere
 $$\bXi \equiv \bxi^2 + i\bxi^3.                                    \m[4-15]$$
  In terms of $\er$ and $\bXi$, the ten Killing vector 4-components
  $\bar\xi^\mu$ or the 2 real + 1 complex components
  $\bxi^\mu\equiv(\bxi^0,\bxi^1,\bXi)$ in $(u,r,x^K)$ coordinates are as
  follows:
 \item{(i)} Time Translations ${\bar\xi}_{(0)}^\mu$:
  $$\bxi^\mu_{(0)}=\delta^\mu_0=(1~,~0~,~0),                        \m[4-16]$$
 \item{(ii)} 3-Space Translations $\{{\bar\xi}_l^\mu \}\equiv\vec{\lam}^\mu$:
  $$\vec{\lam}^\mu=\left(-\er~,~\er~,~{2P^2\over r}\nabla\er   \r), \m[4-17]$$
 \item{(iii)} 3-Space rotations
         $\{\half\epsilon_m^{~~kl}{\bar\xi}_{[kl]}^\mu\}
            \equiv \{\eta^\mu_m\} \equiv\vec{\eta}^\mu$:
  $$\vec{\eta}^\mu=\l(0~,~0~,~2iP^2\nabla\er     \r),               \m[4-18]$$
 \item{(iv)} 3-Spacetime (``Lorentz'') rotations
                $\{ {\bar\xi}_{[0l]}^\mu \}\equiv\vec{\zeta}^\mu$:
  $$\vec{\zeta}^\mu=
       \l( u\er~,~-(r+u)\er~,~-\l(1+{u\over r}\r)2P^2\nabla\er \r). \m[4-19]$$

  These Killing vectors satisfy the Killing  equations
 $$\bar D_\mu \bar\xi_\nu + \bar D_\nu \bar\xi_\mu = 0 ~~~~~,~~~~~
    \bar\xi_\mu = \bar\gmn \bar\xi^\nu                             \m[4-20]$$
  which are very useful in further calculations; using the background metric
  [4-4], the Killing equations [4-20] can be written
   $$\eqalignno{
          \di_1\bxi^0    &=0~~,~~\di_0\bxi^1=\di_1\bxi^1=-\di_0\bxi^0&[4-21]\cr
          \nab(\bxi^0+\bxi^1)&={r^2\over 2P^2}\di_0\bXi              &[4-22]\cr
          \nab\bxi^0         &={r^2\over 2P^2}\di_1\bXi              &[4-23]\cr
          \bxi^1             &=-\half rP^2\l(\bnab{\bXi\over P^2}\r)'&[4-24]\cr
          \nab\bXi           &=0.                                 &[4-25]\cr}$$
  From equation [4-25] applied to ${\vec\eta}^\mu$ defined in [4-18], and from
  equation [4-24] applied to ${\vec\zeta}^\mu$ defined in [4-19], we deduce,
  respectively, the following important identities
 $$\eqalignno{ &\nab(P^2\nab\er)=0                 &   [4-26]  \cr
               &P^2\bnab\nab\er=-\er.              &   [4-27]  \cr}$$

\beginsection 5. Globally Conserved Quantities for Bondi's Asymptotic Metric

  The hypersurface of integration is $x^0=\const$ or $u=u_0$, with boundary
  $\di\Sigma$ at infinity: $x^1=r \goto \infty$. On this boundary the conserved
  quantities ${\cal P}(\bar\xi)$ defined in [2-30] may be written as surface
  integrals:
   $${\cal P}(\bar\xi)={\k \over 8\pi}
      \intl_{\di\Sigma} \hat J^{01}(\bar\xi) dx^2dx^3 ,          \m[5-1]$$
  or in terms of the element of solid angle $d\Omega$
   $$d\Omega=\sin\theta d\theta d\phi={dx^2dx^3 \over 2P^2}      \m[5-2]$$
  eq. [5-1] may be written as
   $${\cal P}(\bar\xi)=\lim_{r\to\infty}
                       \oint_\Omega[P^2\k\hat J^{01}(\bar\xi)]\dom. \m[5-3]$$

  $\k\hat J^{01}(\bar\xi)$ can be deduced from [2-29] or from [2-20],
   $$\k\hat J^{01}(\bar\xi)=\k\hat j^{01}(\bar\xi)
                           -\k\Bar{\hat j^{01}(\bar\xi)}.           \m[5-4]$$
  From [2-15], one obtains
   $$2\k\hat j^{01}(\bxi)
      =\sqrt{-g} ( g^{1\sig}\di_1g_{\sig\lam}\bxi^\lam
                    + \di_1\bxi^1  - g^{1\sig}\di_\sig\bxi^0 )
         + \bxi^0{\hat k}^1 - \bxi^1{\hat k}^0,                     \m[5-5]$$
  which contains ${\hat k}^0$ and ${\hat k}^1$ defined in [2-4].
  The quantity $\Bar{\hat j^{01}}$ is the same as [5-5] written in terms of
  $\bar\gmn$ rather than $\gmn$. Therefore, using $\bar\gmn$ defined in [4-4]
  and eq. [4-21], we have
   $$2\k\Bar{\hat j^{01}(\bar\xi)}=-2\sqrt{-\bar g}\di_0\bxi^0
                                  =-{r^2\over P^2}\di_0\bxi^0;      \m[5-6]$$
  remember that $\Bar{{\hat k}^0}=\Bar{{\hat k}^1}\equiv 0$.

  We substitute the $\gmn$ and $g^{\mu\nu}$ components according to
  [3-2] and [3-5] into [5-5], and calculate $\hat J^{01}$ as given
  in [5-4]. We also use the Killing equations [4-21] and [4-24] to get rid of
  $\bxi^1$. We then place the resulting $\hat J^{01}$ into [5-3] and take the
  limit $r\goto\infty$, to obtain
  the following expression for the integrant of ${\cal P}(\bxi)$:
 $$\eqalignno{\lim_{r\to\infty} P^2 2\k\hat J^{01}(\bxi)=
   &\sig\bsig\di_0\bxi^0
     -\l[ 2\psi_2+2\sig\dot\bsig+P^2\nab\l(f\over P^2\r) \r]'\bxi^0  &[5-7a]\cr
   &+\l[\l( {-\psi_1^*\over P} + \half\bnab(\sig\bsig)
                              - \half{\bsig f^*\over P^2} \r)\bXi\r]'&[5-7b]\cr
   &+\l[P^2\nab \l( {f\over2P^2}\bxi^0
                 -\half{\sig\bsig\over P^2}\bXi^* \r) \r]'           &[5-7c]\cr
   &+ r {1\over 2P^2} [f\bXi]'.                                   &[5-7d]\cr}$$

  The integrant of ${\cal P}(\bxi)$ has been written as a sum of five
  terms. The term [5-7c] gives no contribution to the integral [5-3] because
  --- see [5-2] --- $P^2$ disappears from [5-7c]
  which becomes a pure divergence on a sphere, whose integral is zero. The term
  [5-7d] diverges as $r\goto\infty$. However, the factor of $r$ integrated on
  the sphere is equal to the real part of
 $$\int{1\over 2P^2}f\bXi\dom = {1\over 2\pi}\int{f\over 2P^4}\bXi dx^2dx^3,
                                                                      \m[5-8]$$
  which becomes, using [3-3] and [4-25]
 $$ {1\over 2\pi}\int \nab{\sig\over P^2}\bXi dx^2dx^3
   =-{1\over 2\pi}\int{\sig\over P^2}\nab\bXi dx^2dx^3 = 0 .          \m[5-9]$$
  If it is understood that one takes $r\goto\infty$ after integration, the term
  [5-7d] does not contribute to [5-3]. Thus the integral of
  $P^2 2\k \hat J^{01}$ contains only the factors of $\di_0\bxi^0$ and $\bxi^0$
  in [5-7a], and the term with $\bXi$ in [5-7b]. Further simplifications of
   [5-7] are still possible, but we want to keep it at present in this form.

  We shall now re-write the remaining terms of ${\cal P}(\bxi)$ using the
  Newman and Penrose (1966) ``edth'' operators [see also Newman and Tod 1976],
  to enable comparisons with the formulas found in the literature of the 70's
  and 80's. More precisely, we shall denote by $\de$ the Newman and Penrose
  edth derivative times ${1\over\sqrt 2}$. Thus with formula (3.9) of their
  paper we define:
 $$ \de\eta = {1\over\sqrt 2}\de_{NP66}\eta = P^{1-s}\nab(P^s\eta)   \m[5-10]$$
  where $s$ is the spin weight of $\eta$, $\SW(\eta)=s$.
  The complex conjugate edth derivative $\dez$ [see formula (3.17) in NP66] is
  defined by
 $$ \de^*\eta = P^{1+s}\nab^*(P^{-s}\eta).                          \m[5-11]$$
  As far as our calculation goes, we must know the following spin weights:
 $$\eqalign{
   &\SW(\sig)=2~~,~~\SW(\bsig)=-2~~,~~\SW(\dez\sig)=1~~,~~\SW(\de\bsig)=-1 \cr
   &\SW(\sig\bsig)=0~~,~~\SW(\er)=0~~,~~\SW(\de\er)=1~~,~~\SW(\dez\er)=-1.\cr}
                                                                     \m[5-12]$$
  In terms of [5-10], [5-11] and [5-12], and with [3-3], we can re-write some
  of the quantities appearing in [5-7a,b] as follows:
 $$\eqalignno{
           {f\over 2P}  & = P^3\nab(P^{-2}\bsig) = \de\bsig        &[5-13] \cr
   P^2\nab{f\over 2P^2} & = P^2\nab\l(P^{-1}{f\over 2P}\r)
                           = \de{f\over 2P} = \de^2\bsig           &[5-14] \cr
   {(\sig f)^* \over 2P} & = \bsig\dez\sig                         &[5-15] \cr
   P\nab\er              & = \de\er.                              &[5-16]\cr}$$
  Eq. [4-26] and [4-27] can also be re-written as
 $$\eqalignno{ \de^2\er&=0   &[5-17] \cr      \dez\de\er&=-\er   &[5-18] \cr}$$
  For integration by parts, the following property [equation (3.26) in NP66]
  will be particularly useful
 $$\SW(A)+\SW(B)=-1~~~\Rightarrow~~~
                             \oint(\de A)B\dom=-\oint(\de B)A\dom.   \m[5-19]$$

  With [5-13] to [5-16], the integral [5-3] of the integrant [5-7] may be
  written, taking account of [5-8] and [5-9]
 $$\eqalign{ {\cal P}(\bxi)=
     & \half\oint \sig\bsig\di_0\bxi^0 \dom
       - \oint [\psi_2+\sig\dot\bsig+\de^2\bsig]' \bxi^0 \dom     \cr
     & - \oint \l[ [\psi_1+\sig\de\bsig-\half\de(\sig\bsig)]
                                    {\bXi^*\over 2P} \r]'\dom. \cr} \m[5-20]$$
  Equation [5-20] for ${\cal P}(\bxi)$ has a close ressemblance with the
  standard expressions build on a quite different basis. For comparison see for
  instance Winicour's (1980) formula given in [6-2] below. Equation [5-20] will
  now be simplified further.
  If we look at [4-17] and [4-19], we can see that the $u$-components of the
  Killing vectors $\bxi^0$ for space translations and Lorentz rotations are of
  the form $\bxi^0={\cal F}(u)\er$. For these $\bxi^0$'s, and with
  [5-17], the integral of the $\de^2\bsig$-term in [5-20] is zero; indeed
 $$  -{\cal F}(u)\oint [\de^2\bsig\er]'\dom
   = +{\cal F}(u)\oint [\de\bsig\de\er]'\dom
   = -{\cal F}(u)\oint [\bsig\de^2\er]'\dom = 0.                     \m[5-21]$$
  For the time translation the $u$-component $\bxi_{(0)}^0=1$ (see [4-16]), and
  for this $\bxi^0$ the integral of the $\de^2\bsig$-term in [5-20] is equally
  zero:
 $$ {1\over 4\pi}\oint\de^2\bsig d\Omega
   ={1\over 8\pi}\oint\nab{f\over 2p^2} dx^2dx^3 = 0.                \m[5-22]$$
  The $\de^2\bsig$-term does not contribute to space rotations for which
  $\vec\eta^0$ is zero (see [4-18]).
  Therefore $\de^2\bsig$ may be omitted from the integral [5-20], which reduces
  to
 $$\eqalignno{ {\cal P}(\bxi)=
     & \half\oint \sig\bsig\di_0\bxi^0 \dom             &        [5-23a] \cr
     & - \oint [\psi_2+\sig\dot\bsig]' \bxi^0 \dom      &        [5-23b] \cr
     & - \oint \l[ [\psi_1+\sig\de\bsig-\half\de(\sig\bsig)]
                                    {\bXi^*\over 2P} \r]'\dom. & [5-23c] \cr}$$

  Notice that $\bXi^*=\bxi^2-i\bxi^3$ contributes only through their principal
  parts, with $r\goto\infty$. There is no $\bxi^1$ in [5-23]. If we denote
  the principal parts of the $0,2,3$ components of $\bxi^\mu$ by
  $\bxi^a=(\bxi^0,\bXi_{(r=\infty)})$ $a=0,2,3$  we find, in the notations
  defined in [4-16] to [4-19] and in NU-coordinates that
 $$\eqalignno{   \bxi_{(0)}^a&=(1 ~,~ 0)             &[5-24a] \cr
                   \vec\lam^a&=(-\er ~,~ 0)          &[5-24b] \cr
                   \vec\eta^a&=(0 ~,~ 2iP\de\er)     &[5-24c] \cr
                  \vec\zeta^a&=(u\er ~,~ -2P\de\er). &[5-24d]\cr} $$
  The term $\di_0\bxi^0=0$ except for Lorentz rotations $\vec\zeta^a$, for
  which
        $$\di_0\vec\zeta^a=(\er,0).                                 \m[5-25]$$
  The $\bxi^a$'s are the generators of the Poincar\'e subgroup of the BMS
  transformations at null infinity ($r=\infty$) [see Sachs 1962a, see also
  Newman and Unti 1966].

\beginsection 6. Detailed Comparison with Standard Results

  According to Dray and Streubel (1984), the expressions for
  angular momentum on the cross section $S=\{u=0\}$ given by Tamburino and
  Winicour (1966), Bramson (1975), Lind {\it et al} (1972), Prior (1977),
  Winicour (1968,1980) and Geroch and Winicour (1981) are all of the same form,
  though some of these authors differ by `anomalous' factors of 2 [see below].
  For definiteness we shall compare [5-23] with the explicit expression for
  conservation laws $L_\xi(\Sigma^+)$ given in Winicour (1980) [his
  equation (2.16)]. The comparison is made easier with the following
  redefinitions of Winicour's notations, represented here with an index $W$:
 $$\eqalign{
            &g_{\mu\nu_W}=-\gmn~~{\rm 1968~paper}
                            ~~,~~\psi_{j_W}=2\sqrt 2\psi_j~~j=1,2      \cr
            &\de_{_W}=-\sqrt 2\de~~,~~\sig_{_W}=\sqrt 2\sig
                            ~~,~~\dot\sig_{_W}=2\dot\sig~~
                               {\rm as}~~u_{_W}={u\over\sqrt 2}.  \cr}\m[6-1]$$
  With [6-1], $L_\xi(\Sigma^+)$ can be writen
 $$\eqalignno{  L_\xi(\Sigma^+)=
    &-\oint[(\psi_2+\sig\dot\bsig-\de^{*2}\sig)(\xi^\mu l_\mu)]'\dom &[6-2a]\cr
    &-2\oint[(\psi_1+\sig\de\bsig+\half\de(\sig\bsig))(\xi^\mu m^*_\mu)]'\dom
                                                                  &[6-2b]\cr}$$
  in which
 $$l_\mu=(1~,~0~,~0~,~0)                                            \m[6-3a]$$
  and\footnote*{We are somewhat uncertain about the sign of $m_\mu^*$, and were
  unable to decide.}
 $$m_\mu^*={1\over 2P}(0~,~0~,~1~,~-i);                             \m[6-3b]$$
  $\xi^\mu$ are the dominant parts of the asymptotic symmetry generators.
  Notice therefore that $\xi^\mu m^*_\mu$ in [6-2] is related to the conjugate
  of $\bXi$ defined in [4-15] as follows:
 $$\xi^\mu m^*_\mu = -{1\over 2P}\bXi^*                           \m[6-4]$$

\beginsection {\it (i) Energy and energy flux}

  For time translations (see [5-24a]), we obtain from [5-23] an
  expression for the total energy
 $$E={\cal P}(\bxi_{(0)})=
      -\oint\l[\psi_2+\sig\dot\bsig\r]'{d\Omega\over 4\pi},     \m[6-5]$$
  which is the expression given by Penrose (1964).
  Moreover, using [3-7] to eliminate $\dot\psi_2$ we can calculate the flux
  $dE/du$ of energy in terms of $\sig$:
 $$ {dE\over du} = -\oint \dot\sig\dot\bsig \dom.                    \m[6-6]$$
  This is Bondi's (Bondi, van der Burg \& Metzner 1962) mass loss formula.
  Both $E$ and $dE/du$ are also obtained from [6-2], using [5-22].

\beginsection {\it (ii) Linear momentum and linear momentum flux}

  For space translations (see [5-24b]), [5-23] provides an
  expression for the linear momentum
 $$\P={\cal P}(\vec\lam)= +\oint \l[\psi_2+\sig\dot\bsig\r]' \er\dom, \m[6-7]$$
  Which has also been proposed by Penrose (1964). Using again [3-7] and [4-7]
  the linear momentum flux is
 $$ {d\P\over du} = \oint \dot\sig\dot\bsig \er \dom.                 \m[6-8]$$
  This $d\P/du$ is the BMS invariant flux first given by Sachs (1962b).
   Both $\P$ and $d\P/du$ follow also from [6-2], using [5-21].

\beginsection {\it (iii) Angular momentum and angular momentum flux}

  For spatial rotations in the background, defined by [5-24c], the
  corresponding conserved vector defined in the background is the angular
  momentum $\L$
 $$\L = {\cal P}(\vec\eta)
      =-\oint \l[(\psi_1+\sig\de\bsig)\de^*\er\r]'' \dom.             \m[6-9]$$
  Notice that the term $\half\de(\sig\bsig)\bXi^*/2P$, both in [5-22c] and in
   [6-2], drops out of the integral; indeed, using [5-19], we can write
 $$\half\l[\oint\de(\sig\bsig){\bXi^*\over 2P}\dom\r]'
   =\half\l[\oint\de(\sig\bsig)(-i\dez\er)\dom\r]'
   =\half\l[i\oint (\sig\bsig) \de\dez\er \dom\r]'                   \m[6-10]$$
  which is obviously zero because the {\sl integral} is real.
  As a result, [6-2] gives twice (or perhaps minus
  twice) the value of $\L$ defined by [6-9]. The factor
  2 is the well known anomaly pointed out by Penrose (1982) which makes that
  the Tamburino and Winicour (1966) as well as the Geroch and Winicour (1981)
  definitions do not agree with what one expect in the classical interpretation
  of the quantized weak gravitational field (Gupta 1952). The definitions of
  Streubel (1978), Dray and Streubel (1984), Dray (1985), Shaw (1986) and
  Penrose (1982) do not have the anomalous factor 2. The angular
  momentum flux can be written in the following form using [3-6] and [4-9]
 $${d\L \over du} = -\oint[(\bsig\de\dot\sig-\sig\de\dot\bsig
                                  +2\dot\sig\de\bsig) \dez\er]''\dom.\m[6-11]$$

\beginsection {\it (iv) The mass center initial position}

  For Lorentz rotations in the background defined by [5-24d] and [5-25], The
  conserved vector is
 $$\vec Z = {\cal P}(\vec\zeta)
          =-u\P+\oint\l[(\psi_1+\sig\de\bsig)\de^*\er\r]' \dom,      \m[6-12]$$
  because the term with $\di_0\bxi^0$ in [5-23a] cancels the term with
          $\de_(\sig\bsig)$ in [5-23c]. The flux of $\vec Z$ is
 $$ {d\Z\over du} = - u{d\P\over du}
          -\oint [(\bsig\de\dot\sig+\sig\de\dot\bsig) \dez\er]' \dom.\m[6-13]$$
 Following [2-31], $\Z$ is the expression one expects in the weak field
 approximation.

 The conserved quantity $\Z_W$ deduced from [6-2] is different:
 $$\Z_W=-u\P
    +2\oint\l[(\psi_1+\sig\de\bsig+\half\de|\sig|^2)\de^*\er\r]'\dom;\m[6-14]$$
 remember the sign uncertainty of the integral.

\beginsection 7. Concluding Remarks

  ${\cal P}(\bxi)$ defined in [2-30] is a coordinate independent
  expression. There are ten ${\cal P}(\bxi)$, one for each of the ten Killing
  fields. For Poincar\'e transformations in Minkowski space, the $\bxi$'s and
  therefore also ${\cal P}(\bxi)$'s transform as vectors or tensors like in
  Special Relativity. Bondi's metric is, on the other hand, form invariant for
  supertranslations, [3-9] with $J=1$
  and $\tilde x^K=x^K$
                         $$ u = \tilde u - K(x^L).           \m[7-1] $$
  Such transformations induce changes in the values of the scalar
  functions $\sig$, $\psi_1$ and $\psi_2$. In particular
 $$ \sig(u,x^K) = \tilde\sig(\tilde u,x^K) - \de^2 K.        \m[7-2] $$
  The changes in $d{\cal P}/du$ are obtained with [7-2] from [6-6], [6-8],
  [6-11] and [6-13]. In particular, since
 $$ {\di\sig \over \di u} = {\di\tilde\sig \over \di\tilde u},\m[7-3]$$
  we see that the fluxes of the energy and of the linear momuntum are unchanged
  by supertranslations, except for a re-labelling of $u$ into $\tilde u$; this
  is a well known result.

  We have thus found that the total energy, linear and also the total angular
  momentum at null
  infinity obtained from N\oe ther's theorem and a Lagrangian quadratic in
  first order derivatives, are the same as those considered as standard. The
  constructions of the standard quantities are all, more or less, following the
  general principles outlined in Ashtekar and Winicour (1982). The use of
  Penrose's conformal spaces with the BMS symmetry group and the NP formalism
  is in some ways more appealing and more elegant than the background formalism
  used here. On the other hand, the mapping independant formulation has a
  serious drawback pointed out by Goldberg (1990): globally conserved
  quantities are unrelated to the source of gravity through Einstein's
  equations. While our background metric formalism does not have this defect,
  it must be said that without a mapping rule the formalism remains
  incomplete, and we are stuck like everybody else with supertranslation
  ambiguities. Nevertheless, the fact that eq [2-30] relates ${\cal P}$ to the
  sources of gravitation is of potential value and fills, at least in
  principle, an important gap in the BMS invariant constructions on spheres at
  infinity.

\filbreak
\eject

\beginsection Appendix

 See also KBL96 for detailed calculations.

\beginsection {\it (i)}

 $$\hat\theta^\mu_\nu=\hat T^\mu_\nu - \Bar{\hat T^\mu_\nu}
                      +\txt{1\over 2\k}\hat l^{\rho\sig}\bar R_{\rho\sig}
                                                                \delta^\mu_\nu
                      +\hat t^\mu_\nu ,                 \m[A-1]$$
 in which
  $$ \hat l^{\mu\nu}=\hat g^{\mu\nu} - \Bar{\hat g^{\mu\nu}} ,   \m[A-2]$$
 and
  $$\eqalign{ 2\k {\hat t}^\mu_\nu=
    &{\hat g}^{\rho\sig}\l[ ( \Delta^\lam_{\rho\lam}\Delta^\mu_{\sig\nu}
                       +\Delta^\mu_{\rho\sig}\Delta^\lam_{\lam\nu}
                       -2\Delta^\mu_{\rho\lam}\Delta^\lam_{\sig\nu} )
           -\delta^\mu_\nu( \Delta^\eta_{\rho\sig}\Delta^\lam_{\eta\lam}
                          -\Delta^\eta_{\rho\lam}\Delta^\lam_{\eta\sig}) \r]\cr
     &+{\hat g}^{\mu\lam} ( \Delta^\sig_{\rho\sig}\Delta^\rho_{\lam\nu}
                   -\Delta^\sig_{\lam\sig}\Delta^\rho_{\rho\nu} ).\cr}\m[A-3]$$

\beginsection {\it (ii)}

 $\hat\sig^{\mu[\rho\sig]}$ is the antisymmetric part of
 $\hat\sig^{\mu\rho\sig}$ and
  $$2\k\hat\sig^{\mu\rho\sig}=
      ( g^{\mu\rho} \bar g^{\sig\nu} + g^{\mu\sig} \bar g^{\rho\nu}
       -g^{\mu\nu} \bar g^{\rho\sig} ) \hat\Delta^\lam_{\nu\lam}
    - ( g^{\nu\rho} \bar g^{\sig\lam} + g^{\nu\sig} \bar g^{\rho\lam}
       -g^{\nu\lam} \bar g^{\rho\sig} ) \hat\Delta^\mu_{\nu\lam}.     \m[A-4]$$
 The two terms containing $\bar g^{\rho\sig}$ do not contribute to
 $\hat\sig^{\mu[\rho\sig]}$.

\beginsection {\it (iii)}

 $$\eqalign{ 4\k\zeta^\mu =
       &(Z^\mu_\rho g^{\rho\sig} + g^{\mu\rho} Z^\sig_\rho - g^{\mu\sig} Z)
	                                               \Delta^\lam_{\sig\lam}
        +(g^{\rho\sig} Z - 2g^{\rho\lam} Z^\sig_\lam) \Delta^\mu_{\rho\sig} \cr
       &+l^{\mu\lam} \di_\lam Z
        +l^{\rho\sig} (\bar D^\mu Z_{\rho\sig} -2\bar D_\rho Z^\mu_\sig),\cr}
                                                                      \m[A-5]$$
 in which
  $$Z_{\rho\sig} \equiv \Lix\bar g_{\rho\sig}
                    =   \bar D_\rho \xi_\sig + \bar D_\sig \xi_\rho ~~~~,~~~~
    Z = \bar g^{\rho\sig} Z_{\rho\sig}  ~~~~{\rm and}~~~~
    \xi_\sig = \bar g_{\sig\mu} \xi^\mu.                              \m[A-6]$$

\beginsection References

\baselineskip = 18 pt

{\eightrm
\lft{Arnowitt R, Deser S \& Misner C W 1962.
        \vtop{\hbox{in {\eit Gravitation: An Introduction to Current Research}}
              \hbox{L Witten (New York: Wiley)}}}
\vskip 11 pt
\lft{Ashtekar A \& Winicour J 1982. {\eit J. Math. Phys.} {\ebf 23} 2410}
\lft{Bergmann P G 1949.{\eit Phys. Rev.} {\ebf 75} 680}
\lft{Bergqvist G 1992. {\eit Class. Quantum Grav.} {\ebf 9} 1753}
\lft{Bogoliubov N N \& Shirkov D V 1959.
               \vtop{\hbox{\eit Introduction to the Theory of Quantized Fields}
                     \hbox{(Interscience Publ., New York)}}}
\vskip 11 pt
\lft{Bondi H 1960. {\eit Nature} {\ebf 186} 535}
\lft{Bondi H, van der Burg M G J \& Metzner A W K 1962.
           {\eit Proc. R. Soc. London} {\ebf A269} 21}
\lft{Bramson B D 1975. {\eit Proc. R. Soc. London} {\ebf A341} 463}
\lft{Choquet Bruhat Y 1984.
  \vtop{\hbox{in {\eit Relativity, Groups and Topology II}}
        \hbox{Ed. De Witt C \& Stora R (North Holland, Amsterdam)}}}
\vskip 11 pt
\lft{Chru\'sciel P 1986.
   \vtop{\hbox{in {\eit Topological Properties and Global Structure
                        of Spacetime}}
         \hbox{Ed. P G Bergmann \& V De Sabatta, Series B: Physics {\ebf 138}
               Nato ASI series}
         \hbox{(North Holland, Amsterdam)}}}
\vskip 11 pt
\lft{Cornish F H J 1964.  {\eit Proc. R. Soc. London} {\ebf A282} 358, 372}
\lft{Cresswell \& Zimmerman 1986. {\eit Class. Quantum Grav.} {\ebf 3} 1221}
\lft{Dray T 1985. {\eit Class. Quantum Grav.} {\ebf 2} L7}
\lft{Dray T \& Streubel M 1984. {\eit Class. Quantum Grav.} {\ebf 1} 15}
\lft{Eisenhart L P 1949. {\eit Riemannian Geometry}
                                             (Prinston University Press) p. 85}
\lft{Freud P 1939. {\eit Ann. Math. J.} {\ebf 40} 417}
\lft{Geroch R 1970. {\eit J. Math. Phys.} {\ebf 11} 2580}
\lft{Geroch R 1976. in {\eit Asymptotic Structure of Spacetime} Ed. Esposito P
                                          and Witten L (Plenum, New York) p. 1}
\lft{Geroch R \& Winicour J 1981. {\eit J. Math. Phys.} {\ebf 22} 803}
\lft{Goldberg J N 1990. {\eit Phys. Rev. D} {\ebf 41} 410}
\lft{Gupta S N 1952. {\eit Proc. Phys. Soc.} {\ebf 65,3-A} 161}
\lft{Katz J 1985. {\eit Class. Quantum Grav.} {\ebf 2} 423}
\lft{Katz J 1996.
   \vtop{\hbox{ in {\eit Gravitational Dynamics} Ed. Lahav O, Terlevich E
                \& Terlevich R J }
         \hbox{ (Cambridge Univ. Press) p. 193}}}
\vskip 11 pt
\lft{Katz J, Bi{\u c}{\'a}k J \& Lynden-Bell D 1996.
   \vtop{\hbox{``Relativistic Conservation Laws and Integral Constraints
                                                                in Cosmology''}
         \hbox{(Pre-print)}}}
\vskip 11 pt
\lft{Katz J \& Ori A 1990. {\eit Class. Quantum Grav.} {\ebf 7} 787}
\lft{Komar A 1959. {\eit Phys. Rev.} {\ebf 113} 934}
\lft{Landau L D \& Lifshitz E M 1951. {\eit The Classical Theory of Fields}
                                                       (Addison Wesley) p. 317}
\lft{Lerer D 1996.
   \vtop{\hbox{ {\eit Angular Momentum of Radiating Spacetimes} MSc thesis,
                 The Racah Institute of Physics,}
         \hbox{ Hebrew University of Jerusalem (unpublished) Appendix}}}
\vskip 11 pt
\lft{Lind R W, Messmer J and Newman E T 1972. {\eit J. Math. Phys.}
                                                                {\ebf 13} 1884}
\lft{Misner C W, Thorne K S and Wheeler J A 1973. {\eit Gravitation} (Freeman,
                                                         San Francisco) p. 452}
\lft{M\o ller C 1972. {\eit The Theory of Relativity} 2nd Ed. (Clarendon,
                                                                       Oxford)}
\lft{Newman E T \& Penrose R 1962. {\eit J. Math. Phys.} {\ebf 3} 566}
\lft{Newman E T \& Penrose R 1966. {\eit J. Math. Phys.} {\ebf 7} 863}
\lft{Newman E T \& Tod K P 1976.
    \vtop{\hbox{In {\eit Asymptotic Structure of Spacetime}}
          \hbox{Ed. Esposito P and Witten L (Plenum, New York) p. 229}}}
\vskip 11 pt
\lft{Newman E T \& Unti T W J 1962. {\eit J. Math. Phys.} {\ebf 3} 891}
\lft{Papapetrou A 1974. {\eit Lectures on General Relativity}
                        (Reidel Pub., Dordrecht) p. 105, 133}
\lft{Penrose R 1964. in {\eit Relativity, Groups and Topology} (Gordon and
                                                   Breach, New York) p. 565}
\lft{Penrose R 1965. {\eit Phys. Rev. Lett.} {\ebf 14} 57}
\lft{Penrose R 1982. {\eit Proc. R. Soc. London} {\ebf A381} 53}
\lft{Poincar\'e H 1904.
   \vtop{\hbox{{\eit La Science et L'Hypoth\`ese} (Flamarion, Paris).
                             Extracts are from the 1952 Dover Publications}
         \hbox{translation entitled {\eit Science and Hypothesis} p. 121}}}
\vskip 11 pt
\lft{Prior C R 1977. {\eit Proc. R. Soc. London} {\ebf A354} 379}
\lft{Rosen N 1940. {\eit Phys. Rev.} {\ebf 57} 147, 150}
\lft{Rosen N 1958. {\eit Phys. Rev.} {\ebf 110} 291}
\lft{Rosen N 1963. {\eit Ann. Phys. (NY)} {\ebf 22} 1}
\lft{Sachs R K 1962a. {\eit Phys. Rev.} {\ebf 128} 2851}
\lft{Sachs R K 1962b. {\eit Proc. R. Soc. London} {\ebf A270} 103}
\lft{Schmidt B 1993. in {\eit Advances in Gravitation and Cosmology} (Wiley
                                        Eastern Ltd, New Delhi) p. 81}
\lft{Schweber S S, Bethe H A \& de Hoffmann F 1956.
    \vtop{\hbox{{\eit Mesons and fields} Vol I}
          \hbox{(Row Peterson, New York) appendix B, p. 415}}}
\vskip 11 pt
\lft{Shaw W T 1986. {\eit Class. Quantum Grav.} {\ebf 1} L33}
\lft{Sorkin R D 1988.
  \vtop{\hbox{in {\eit Mathematics and General Relativity} Ed. Isenberg J W}
   \hbox{Vol. {\ebf 71} in the AMS's Contemporary Mathematics series, p. 23}}}
\vskip 11 pt
\lft{Streubel 1978. {\eit Gen. Rel. Grav.} {\ebf 9} 551}
\lft{Synge J L 1964. in {\eit Relativity: The General Theory} (North Holand,
                                                             Amsterdam) p. 249}
\lft{Tamburino L \& Winicour J 1966. {\eit Phys. Rev.} {\ebf 150} 1039}
\lft{Tolman R C 1934. {\eit Relativity, Thermodynamics and Cosmology}
                      (Dover edition 1987) p.233}
\lft{Thorne K S 1980. {\eit Rev. Mod. Phys.} {\ebf 52} 299}
\lft{Weinberg S 1972.
  \vtop{\hbox{\eit Gravitation and Cosmology:
                             Principles and Applications of General Relativity}
        \hbox{(New York: Wiley) p. 380}}}
\vskip 11 pt
\lft{Winicour J 1968. {\eit J. Math. Phys.} {\ebf 9} 861}
\lft{Winicour J 1980. in {\eit General relativity and Gravitation} vol 2,
                                           ed. A Held (New York: Plenum) p. 71}
\lft{Wald R M 1984. {\eit General Relativity} (University of Chicago) p. 271} }

\end